\documentclass{jpp}
\usepackage{amsmath}
\usepackage{graphicx}
\usepackage{epstopdf, epsfig}
\usepackage{hyperref}
\usepackage{bigints}
\usepackage{bm}

\usepackage{color}

\DeclareRobustCommand{\vect}[1]{\bm{#1}}
\newcommand{\bfalpha}{\bm \alpha}
\newcommand{\bfbeta}{\bm \beta}
\newcommand{\bfGamma}{\bm \Gamma}

\newcommand{\bfb}{\mbox{\boldmath$b$}}
\newcommand{\bfh}{\bm h}
\newcommand{\bfk}{\mbox{\boldmath$k$}}
\newcommand{\bfu}{\mbox{\boldmath$u$}}
\newcommand{\bfv}{\bm v}
\newcommand{\bfx}{\mbox{\boldmath$x$}}
\newcommand{\bfzhat}{\bm {\hat z}}

\newcommand{\bfA}{\bm A}
\newcommand{\bfB}{\bm B}
\newcommand{\bfF}{\bm F}
\newcommand{\bfG}{\bm G}
\newcommand{\bfH}{\bm H}
\newcommand{\bfU}{\bm U}
\newcommand{\bfV}{\bm V}

\newcommand{\bmnabla}{\bm \nabla}

\newcommand{\dk}{{\mathrm{d}}k}
\newcommand{\domega}{{\mathrm{d}}\omega}
\newcommand{\dV}{{\mathrm{d}}V}

\newcommand{\calE}{\mbox{$\cal E$}}
\newcommand{\bfcalE}{\mbox{\boldmath$\cal E$}}

\newcommand{\la}{\langle}
\newcommand{\ra}{\rangle}

\newcommand\pt[1]{\frac{\partial#1}{\partial t}}
\newcommand\pd[2]{\frac{\partial#1}{\partial#2}}

\shorttitle{Mean Field Electrodynamics: Triumphs and Tribulations}
\shortauthor{D. W. Hughes}

\title{Mean Field Electrodynamics: Triumphs and Tribulations}

\author{David W. Hughes\aff{1}
  \corresp{\email{d.w.hughes@leeds.ac.uk}}}

\affiliation{\aff{1}Department of Applied Mathematics, University of Leeds, Leeds LS2 9JT, UK}

\begin{document}

\maketitle

\begin{abstract}
The theory of mean field electrodynamics, now celebrating its fiftieth birthday, has had a profound influence on our modelling of cosmical dynamos, greatly enhancing our understanding of how such dynamos may operate. Here I discuss some of its undoubted triumphs, but also some of the problems that can arise in a mean field approach to dynamos in fluids (or plasmas) that are both highly turbulent and also extremely good electrical conductors, as found in all astrophysical settings.
\end{abstract}

\section{The Need for a Mean Field Dynamo Theory}

The origins of dynamo theory, and indeed one might argue the origins of magnetohydrodynamics (MHD) itself, might be traced to the short, but hugely influential paper by \citet{Larmor_1919}, entitled `How could a rotating body such as the Sun become a magnet?', in which he postulated that the swirling motions inside stars could maintain a magnetic field. Theoretical progress following this paper was, however, not swift, and indeed the first concrete result in what we would now call dynamo theory was a negative one --- the celebrated anti-dynamo theorem of \citet{Cowling_1933}, showing that an axisymmetric magnetic field could not be maintained by dynamo action. The first example of a working dynamo did not come until a quarter of a century later, with the paper by \cite{Herzenberg_1958} showing how a magnetic field could be maintained by two widely separated spherical rotors with inclined axes of rotation in an electrically conducting fluid otherwise at rest.

One of the most important and long-standing issues in astrophysical fluid dynamics is to explain the generation of global-scale magnetic fields; i.e.\ magnetic fields with a significant components on the scale of the body in question. In broad terms, one might think of a global-scale dynamo as a mechanism by which toroidal field is created from poloidal field whilst, conversely and simultaneously, the poloidal field is regenerated from the toroidal field. Whereas the winding up of poloidal field into toroidal field can be readily explained as a natural consequence of a differentially rotating flow, essentially by Alfv\'en's theorem, the closing of the `dynamo loop' (i.e.\ the regeneration of  poloidal from toroidal field) is much less straightforward. It is here that the power of mean field electrodynamics really comes into its own --- what is problematic in the full (unaveraged) MHD equations emerges naturally in their mean field counterparts. (Herzenberg's model was ingenious, but his choice of flow cleverly sidestepped this issue.)

The means by which poloidal field could be regenerated from toroidal field in a turbulent flow was first addressed by \cite{Parker_1955}, in one of many striking contributions to astrophysics. Although different in style to later formulations of mean field electrodynamics, this can be recognised as the first paper on what we would now call mean field dynamo theory. Parker's crucial insight was to recognise that small-scale cyclonic motions could raise and twist toroidal field, and that the subsequent coalescence of the field loops thus formed would lead to a large-scale poloidal component of the magnetic field. This led to the vital new ingredient of a source term for the poloidal field and, consequently, to the development of what we would now call an $\alpha \omega$-dynamo, with its associated propagating dynamo waves.

The idea of encapsulating the large-scale influences of small-scale interactions was developed into a beautifully elegant theory by Steenbeck, Krause and R\"adler in a series of papers in the 1960s; these were translated into English by \cite{RS_1971}, and the ideas contained therein were elucidated and developed further in the monographs by \cite{Moffatt_1978} and \cite{KR_1980}. It is on these papers, together with that of \cite{Parker_1955}, that the edifice of mean field electrodynamics has been built over the past fifty years, and which has formed the framework for investigating dynamo action in astrophysical bodies.

In this paper I shall discuss some of the most important results that have arisen from mean field dynamo theory, but shall also point out some of the difficulties encountered in applying the theory to the high conductivity, turbulent regime applicable in astrophysics. In this necessarily somewhat brief review, I shall concentrate on the formulation and fundamental aspects of the theory; it is though, even within these confines, far from comprehensive. Furthermore, over the years, certain aspects of mean field electrodynamics have proved quite controversial; some of the views presented here are certainly subjective, and not all are universally accepted.

\section{Mathematical Formulation of Mean Field Electrodynamics}\label{sec:formulation}

\smallskip

\subsection{Large- and small-scale dynamos}

The aim of mean field electrodynamics is to address what might be termed \textit{large-scale} dynamos. Before
considering the mathematical details of mean field dynamo theory, it is therefore worth discussing the classification of dynamos in terms of their spatial scale. Although any such classification is somewhat imprecise, it does nonetheless highlight some important considerations. Let us first consider the case where there is a fluid velocity, either laminar or turbulent, with a single well-defined characteristic length scale, $\ell_o$ say. Then, in rather general terms, we might categorise a large-scale dynamo as one in which there is a sizeable fraction of the magnetic energy at scales very large compared with $\ell_0$. A more precise, and more satisfactory, requirement for a dynamo to be large scale would be that the spectrum of magnetic energy has a local maximum at large scales (together with another at the scale of the flow). A small-scale dynamo may be defined as one for which the significant scales of the field are comparable with or smaller than $\ell_0$.

A small-scale dynamo can be unequivocally defined, at least computationally or experimentally: if amplification of magnetic energy is found in a domain of size $O(\ell_0)$ then the flow is acting as a small-scale dynamo. Things become less clear in extended domains. If the domain of a small-scale dynamo is extended, then the magnetic energy will, almost certainly, possess some large-scale component; however, unless this is pronounced, it seems reasonable to categorise such a dynamo as essentially small scale. The strictest definition of a large-scale dynamo would be one that succeeds in a large enough domain, but fails when the domain size is reduced to $O(\ell_0)$.

For naturally occurring dynamos in astrophysical bodies, making a clear designation of a dynamo as large scale is more problematic. Not only is the domain size fixed, but the flows typically will not possess a single characteristic scale. For example, if we consider the solar magnetic field, then it is certainly true that it has a significant global-scale component, manifested through its appearance at the surface as active regions; however, it is also the case that the Sun has a strong global-scale flow in the form of its differential rotation. Thus, with both field and flow on the largest scale available, it is hard to go beyond saying that this is a global-scale or system-size dynamo.

\subsection{Deriving the mean field induction equation}

Although there are important and controversial questions concerning the nonlinear (dynamical) aspects of mean field electrodynamics, which we shall discuss later, it is formally a linear (kinematic) theory; we shall therefore first consider its formulation via only the induction equation. Thus, in standard notation, under the simplifying assumption that the magnetic diffusivity $\eta$ is uniform, we consider the equation
\begin{equation}
\pt{\bfB}= \bnabla \times (\bfU \times \bfB) + \eta \nabla^2 \bfB ,
\label{eq:ind}
\end{equation}
where it is assumed that there is no influence of the magnetic field on the velocity. The exposition below follows closely that of \cite{Moffatt_1978}, in which further details can be found.

The underlying assumption of mean field theory is that there is some sort of scale separation between large and small scales, and that one then studies the evolution of averaged (or mean) quantities, where the average (which we shall denote by $\langle \cdot \rangle$) is taken over some intermediate scale. Here we shall consider the separation to be in spatial scales, which is the most natural framework for applications of mean field theory to astrophysical bodies. The velocity and magnetic field may then be expressed in terms of large- and small-scale components (or mean and fluctuating components) as
\begin{equation}
\bfU = \bfU_0 + \bfu, \qquad
\bfB = \bfB_0 + \bfb,
\end{equation}
with $\la \bfb \ra = \la \bfu \ra =0$.

The mean induction equation then takes the form
\begin{equation}
\pt{\bfB_0}= \bnabla \times (\bfU_0\times \bfB_0) + \bnabla \times \bfcalE + \eta \nabla^2 \bfB_0,
\label{eq:mean_ind}
\end{equation}
where $\bfcalE$, the mean electromotive force (e.m.f.), is defined by
\begin{equation}
\bfcalE=\la \bfu \times \bfb \ra,
\end{equation}
thus representing the projection onto the large scale of the interactions between the small-scale velocity and small-scale magnetic field. It is the presence of the term involving the e.m.f.\ that sets the mean induction equation apart from its unaveraged counterpart. Since here we are interested particularly in the formulation and interpretation of mean field electrodynamics, we shall from now on ignore the influence of the mean flow, setting $\bfU_0=0$. Interactions between the mean flow and mean magnetic field, although extremely important, as discussed in the introduction, are not germane to considerations of the mean e.m.f. Hence \eqref{eq:mean_ind} simplifies to
\begin{equation}
\pt{\bfB_0} =  \bnabla \times \bfcalE + \eta \nabla^2 \bfB_0.
\label{eq:mean_ind_nomeanflow}
\end{equation}

To make progress with the mean field approach, via equation~\eqref{eq:mean_ind_nomeanflow}, one has to express the mean e.m.f.\ in terms of only mean quantities --- the closure problem of mean field MHD. The standard approach is to consider the evolution equation for the fluctuating field, formed by subtracting \eqref{eq:mean_ind_nomeanflow} from \eqref{eq:ind}:
\begin{equation}
\pt{\bfb} =  \bnabla \times (\bfu \times \bfB_0) + \bnabla \times \bfG  + \eta \nabla^2 \bfb,
\label{eq:fluct}
\end{equation}
where $\bfG= \bfu \times \bfb - \la \bfu \times \bfb \ra$. Formally we may write \eqref{eq:fluct} as
\begin{equation}
{\cal L} (\bfb) = \bnabla \times (\bfu \times \bfB_0),
\label{eq:fluct2}
\end{equation}
where $\cal L$ is a linear operator. It is at this stage, in seeking to solve \eqref{eq:fluct2} for $\bfb$, that we encounter the first potential difficulty, one to which we shall return in \S\,\ref{subsec:ssda}, namely whether there are non-decaying solutions to ${\cal L}(\bfb) = 0$. However, if we make the assumption that all solutions of the homogeneous equation ${\cal L} (\bfb)=0$  decay in time, then we can regard the right hand side of \eqref{eq:fluct2} as a source term for the fluctuating field. In this case, the linearity between $\bfb$ and $\bfB_0$, and hence that between $\bfcalE$ and $\bfB_0$, suggests an expansion of $\bfcalE$ in terms of $\bfB_0$ and its derivatives. This is usually written as 
\begin{equation}
\calE_i = \alpha_{ij} B_{0j} + \beta_{ijk} \pd{B_{0j}}{x_k} + \cdots,
\label{eq:emf}
\end{equation}
where the coefficients $\alpha_{ij}$, $\beta_{ijk}$, etc.\ are pseudo-tensors governed by the properties of the velocity field $\bfu(\bfx,t)$ (pseudo-tensors because $\bfcalE$ is a polar vector whereas $\bfB_0$ is axial). Substituting \eqref{eq:emf} into \eqref{eq:mean_ind} yields the mean induction equation for the evolution of $\bfB_0$, namely
\begin{equation}
\pt{B_{0i}} = \epsilon_{ijk} \frac{\partial}{\partial x_j} \left( \alpha_{kl} B_{0l} + \beta_{klm} \pd{B_{0l}}{x_m} + \cdots \right) + \eta \nabla^2 B_{0i}.
\label{eq:mean_ind2}
\end{equation}

It should be noted that in the ansatz~\eqref{eq:emf}, and hence also in the mean induction equation~\eqref{eq:mean_ind2}, although higher order terms are suggested, consideration is only ever given to the first two terms in the expansion; indeed, inclusion of higher order terms would lead to spatial derivatives of higher than second order in the mean induction equation, thus rendering it of a very different mathematical character to the unaveraged induction equation.

For general flows, lacking any symmetry, the $\bfalpha$ and $\bfbeta$ tensors can lead to a wide variety of quite complicated effects \citep[see, for example,][]{KR_1980, Roberts_1994} In order to get to the very essence of the theory, it is though often instructive to consider the simplified case in which the flow is isotropic and in which the mean field tensors therefore simplify to $ \alpha_{ij} = \alpha \delta_{ij}$, $\beta_{ijk} = \beta \epsilon_{ijk}$, where $\alpha$ is a pseudo-scalar and $\beta$ is a pure tensor. With the further simplifying assumption that $\alpha$ and $\beta$ are constants (i.e.\ not dependent on the large spatial scale), equation~\eqref{eq:mean_ind2} simplifies to
\begin{equation}
\pt{\bfB_0} = \alpha \bnabla \times \bfB_0 + \beta \nabla^2 \bfB_0,
\label{eq:mean_ind3}
\end{equation}
assuming $\beta \gg \eta$. Whereas, in this simplest formulation, $\beta$ is clearly an additional, turbulent, contribution to the magnetic diffusivity, the term involving $\alpha$ (the dynamo `$\alpha$-effect') is of a completely different character to the induction term in the unaveraged equation~\eqref{eq:ind}; the mean e.m.f.\ is parallel to the mean field, whereas the e.m.f.\ is orthogonal to the magnetic field. From \eqref{eq:mean_ind3}, the growth rate $p$ of a long wavelength perturbation with wavenumber $k$ is then given by
\begin{equation}
p \sim \alpha k - \beta k^2.
\label{eq:grate}
\end{equation}

The fact that $\alpha$ in \eqref{eq:mean_ind3} is a pseudo-scalar allows us to make an extremely powerful statement about mean field generation at this early stage, without any detailed calculations. If the turbulence lacks reflectional symmetry (or handedness) then its statistical properties, and hence $\alpha$, must remain unchanged on switching from right-handed to left-handed axes; however, by definition, a pseudo-scalar must change sign under this switch. Thus in reflectionally symmetric flows, $\alpha$ must vanish. The simplest measure of the lack of reflectional symmetry of a flow (or, equivalently, of its `handedness') is the (kinetic) helicity, defined by
\begin{equation}
{\cal H} = \la \bfu \cdot \bmnabla \times \bfu \ra.
\end{equation}
It is thus easy to see, without looking too far, why helicity plays such a dominant role in mean field electrodynamics.

Finally in this sub-section we note that while the derivation of the mean field induction equation~\eqref{eq:mean_ind2} is a consequence of an assumed separation of spatial scales, one might also consider the problem of dynamo action in a flow that has two very different \textit{temporal} scales. Interestingly, the mean induction equation, now depicting the evolution of the magnetic field on long timescales, takes a different form, as shown by \cite{HL_2011} and \cite{Vladimirov_2012}. Rather than the new mean field contribution being an $\alpha$-effect, it instead takes the form of an additional Stokes drift velocity. This is an interesting and potentially important branch of mean field dynamo theory, which to date has been relatively unexplored.

\subsection{Calculating the mean field tensors} \label{sec:cmft}

In order to make use of equation~\eqref{eq:mean_ind2} one needs to be able to calculate the tensors $\alpha_{ij}$ and $\beta_{ijk}$. Furthermore, for the mean field theory to be of practical value, this needs to be done in a way that does not involve solving the full dynamo problem. Let us first consider the determination of the $\bfalpha$ tensor. The key thing to note is that in the determination of $\alpha_{ij}$ by \eqref{eq:emf}, the large-scale field $\bfB_0$ can be taken as \textit{uniform}. The idea then is to impose a uniform magnetic field $\bfB_0$ (still kinematic for the time being), to calculate, either analytically or numerically, the resulting e.m.f.\ and then to determine $\alpha_{ij}$ from the relation
\begin{equation}
\calE_i = \alpha_{ij} B_{0j}.
\end{equation}
To determine all nine components of $\alpha_{ij}$ requires consideration of three independent imposed fields $\bfB_0$. This all sounds straightforward, and sometimes it is, but there are a couple of important subtleties lurking beneath the surface: one is whether the prescription described does indeed even lead to a well-defined value of $\alpha_{ij}$; the other is whether the calculated $\alpha_{ij}$ is the critical feature in determining the growth of any subsequent dynamo. We shall explore both of these issues in subsequent sections.

In principle, the calculation of $\beta_{ijk}$ proceeds in a similar fashion, following the imposition of fields of uniform gradient; there is here though one further issue, which we address immediately below.

\subsection{An extended expansion for the e.m.f.}

We note that the expansion~\eqref{eq:emf} contains spatial, but not temporal derivatives of $\bfB_0$. Although, using \eqref{eq:mean_ind}, it might be argued that one could formally substitute for time derivatives of $\bfB_0$ in terms of spatial derivatives, expression~\eqref{eq:emf} does omit a crucial part of the diffusion tensor. The problem arises because of the way in which $\beta_{ijk}$ is calculated --- namely from a spatially-dependent but time-independent mean field, whereas, in reality, a mean field with spatial dependence will also vary with time. To clarify this, we may, following \cite{HP_2010}, instead expand the e.m.f.\ as
\begin{equation}
\calE_i = \alpha_{ij} B_{0j} + \Gamma_{ij} \pt{B_{0j}} + \beta_{ijk} \pd{B_{0j}}{x_k} + \cdots,
\label{eq:emf_t}
\end{equation}
where the tensors $\alpha_{ij}$ and $\beta_{ijk}$ are identical to those in expression~\eqref{eq:emf}. Assuming, plausibly, that the expression for the e.m.f.\ is a rapidly convergent series, we may, at \textit{this} stage, back substitute for the time derivative of $B_0$; on retaining just the leading order terms, this gives
\begin{equation}
\pt{B_{0i}}= \epsilon_{ijk} \frac{\partial}{\partial x_j} \left( \alpha_{km}B_{0m}+ \Gamma_{km} \epsilon_{mpq} \frac{\partial}{\partial x_p} (\alpha_{qr}B_{0r}) + \beta_{kmn} \pd{B_{0m}}{x_n} \right).
\end{equation}
In the simplest case when the tensors $\bfalpha$, $\bfbeta$ and $\bfGamma$ are constants (more generally, they could be functions of the slow spatial variation), it can be seen that the coefficient of the second order spatial derivative term is not $\beta_{ijk}$, but is instead
\begin{equation}
 \epsilon_{mkq} \alpha_{qj} \Gamma_{im}+ \beta_{ijk}. 
\end{equation}
In the light of the prescription described in \S\,\ref{sec:cmft}, it should be noted that the first component in this expression is simply unattainable by starting from the expansion \eqref{eq:emf} and calculating $\beta_{ijk}$ by considering steady mean currents. To determine $\Gamma_{ij}$ we should consider a uniform magnetic field that increases linearly in time, thus precluding contributions from all terms except the first two in the expansion~\eqref{eq:emf_t}. \cite{HP_2010} looked at this question in some detail, showing how at low values of the magnetic Reynolds number $Rm$ the new term is dominated by the traditional $\beta$ diffusion tensor, but that at higher $Rm$ it can itself become the dominant term.

\section{The Kinematic Regime}\label{sec:KinReg}

\smallskip

\subsection{When it all works beautifully}

Formally, the way forward for obtaining expressions for $\bfalpha$ and $\bfbeta$ is clear. Solution of the fluctuating induction equation~\eqref{eq:fluct} gives $\bfb$ in terms of the flow $\bfu$ and mean field $\bfB_0$; the mean e.m.f.\ can then be determined in terms of $\bfu$ and $\bfB_0$ and thus the mean field tensors determined. However, solution of \eqref{eq:fluct} is made problematic by the presence of the $\bfG$ term involving the fluctuating e.m.f.; it is therefore instructive to consider circumstances in which we might be rid of this troublesome term. There are two such cases: one is when the magnetic Reynolds number $Rm$ is very small; the other is when the correlation time of the flow is assumed to be so short that one may employ what is known as the `short sudden approximation'.

If $Rm \ll 1$ then the term involving $\bfG$ is formally $O(Rm)$ smaller than the diffusive term and can be neglected \citep[see][]{Moffatt_1978}; this is often referred to as the \textit{first order smoothing approximation}. It is then possible to solve for $\bfb$, via a Fourier transform, and hence express $\alpha_{ij}$ and $\beta_{ijk}$ in terms of the spectrum tensor of the velocity field $\bfu(\bfx, t)$. For the simplest case of isotropic turbulence this leads to the results
\begin{equation}
\alpha = - \frac{\eta}{3} \bigintsss \! \! \! \! \bigintsss \frac{k^2 F(k, \omega)}{\omega^2 + \eta^2 k^4} \, \dk \, \domega, \qquad
\beta =  - \frac{2}{3} \eta \bigintsss \! \! \! \! \bigintsss \frac{k^2 E(k, \omega)}{\omega^2 + \eta^2 k^4} \, \dk \, \domega,
\label{eq:coeffs_low_Rm}
\end{equation}
where the integrals are over wavenumber and frequency \citep{Moffatt_1978}; $E(k,\omega)$ and $F(k,\omega)$ are, respectively, the energy and helicity spectrum functions.

In the short sudden approximation \citep{Parker_1955, KR_1980}, it is assumed that correlations between $\bfu$ and $\bfb$ are so short lived that they can be neglected; this may be regarded as the case of very small Strouhal number, defined by $S= U \tau_c/L$, where $U$ and $L$ are representative values of the velocity and length scale of the turbulence and $\tau_c$ is the correlation time. Furthermore, it is assumed that the electrical conductivity is high ($Rm\gg 1$) and hence that the diffusive term can also be ignored. Under these fairly drastic assumptions, equation~\eqref{eq:fluct} becomes
\begin{equation}
\pt{\bfb} \approx  \bnabla \times (\bfu \times \bfB_0) ,
\label{eq:fluct_ss}
\end{equation}
the solution of which is then approximated by 
\begin{equation}
\bfb \approx \tau_c \bnabla \times (\bfu \times \bfB_0).
\label{eq:b_ss}
\end{equation}
From this stage we can readily evaluate the mean field tensors; for isotropic turbulence we have the well-known results
\begin{equation}
\alpha = - \frac{\tau_c}{3} \la \bfu \cdot \bnabla \times \bfu \ra, \qquad \ 
\beta = -\frac{\tau_c}{3} \la \bfu^2 \ra .
\label{eq:coeffs_ss}
\end{equation}
Expressions~\eqref{eq:coeffs_low_Rm} and \eqref{eq:coeffs_ss} both exhibit a strong link between $\alpha$ and the helicity; indeed, under the short sudden approximation, they are directly correlated. Furthermore, these expressions show how a complex magnetohydrodynamic problem, namely the generation of a mean magnetic field, can be simplified enormously to one where the $\alpha$-effect can be related to a single characteristic of the flow. These ideas have led to the notion that helicity (a natural consequence of flow in a rotating body) implies a healthy $\alpha$-effect, which, in turn, leads to a significant mean magnetic field.  However, both \eqref{eq:coeffs_low_Rm} and \eqref{eq:coeffs_ss} are derived under assumptions that are not applicable in astrophysical turbulence, which is characterised by extremely high values of $Rm$ (though discarding the diffusive term is always risky) but with $O(1)$ values of $S$. 

\subsection{More complicated behaviour: reinstating $\vect{G}$}

\begin{figure}
  \centerline{\includegraphics[width=0.85\hsize]{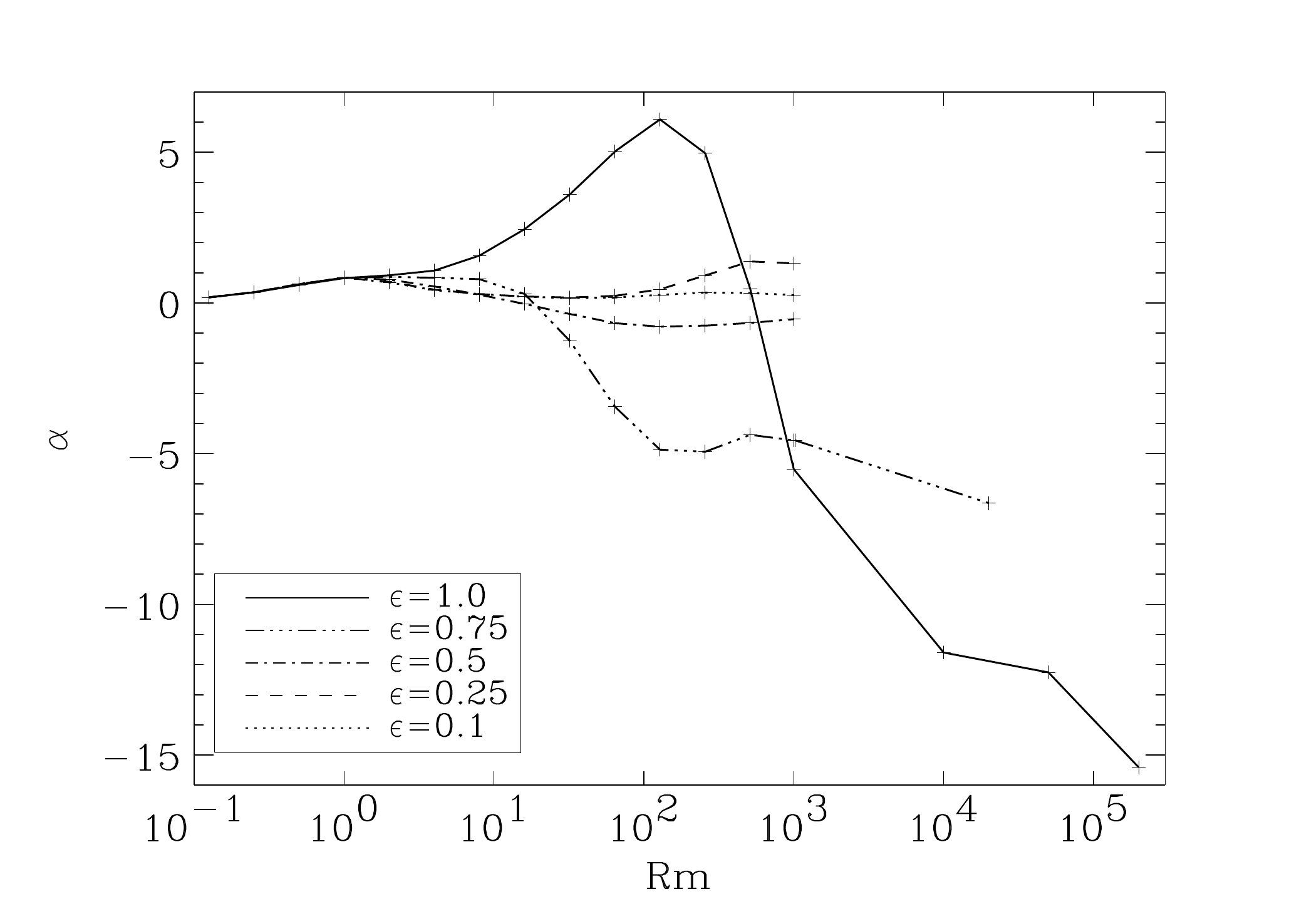}}
  \caption{$\alpha$ as a function of $Rm$ for the flows \eqref{eq:flows} (from \cite{CHT_2006}).}
\label{fig:CHT}
\end{figure}

As we have seen, neglecting $\bfG$ in \eqref{eq:fluct} allows us to obtain concise expressions for the mean field coefficients encapsulating characteristics of the flow. Unfortunately, life is typically not that simple; correlations between $\bfu$ and $\bfb$ do matter, $\bfG$ needs to be included, and the results are not so straightforward. This may be seen in a conceptually straightforward manner by considering a prescribed flow with a given energy and helicity and then calculating how $\alpha$, for example, varies with $Rm$. Following the prescription outlined in \S\,\ref{sec:cmft}, \cite{CHT_2006} considered the family of $z$-independent flows
\begin{equation}
\bfu=(\partial_y \psi, - \partial_x \psi, \psi), \quad \textrm{with} \quad
\psi(x,y,t) = \sqrt{\frac{3}{2}} \left( \cos(x+ \epsilon \cos t) + \sin(y+ \epsilon \sin t) \right),
\label{eq:flows}
\end{equation}
and calculated the e.m.f., and hence $\alpha$, following the imposition of a kinematic uniform magnetic field in the $xy$-plane. The flows \eqref{eq:flows} are maximally helical (i.e.\ $\bfu$ is parallel to $\bnabla \times \bfu$); the steady flow with $\epsilon=0$ is that first introduced by \cite{Roberts_1972} in one of the early calculations of the $\alpha$-effect; flows with $\epsilon \ne 0$ are chaotic, the $\epsilon=1$ case being studied by \cite{GP_1992} as a candidate for fast dynamo action. Note that here we are considering a purely two-dimensional problem; there is no dynamo action and hence the measured e.m.f.\ is proportional to the imposed field; i.e., it provides a `clean' calculation of the $\alpha$-effect. Figure~\ref{fig:CHT} shows $\alpha$ as a function of $Rm$. What is most striking is that there is no longer any clear link between $\alpha$ and helicity; indeed, for a fixed flow, $\alpha$ can change sign as $Rm$ varies, leading to special isolated values of $Rm$ for which there is no $\alpha$-effect whatsoever, even though the flow is maximally helical!

\subsection{Averaging: a question of coherence}

\begin{figure}
  \centerline{\includegraphics[width=0.85\hsize]{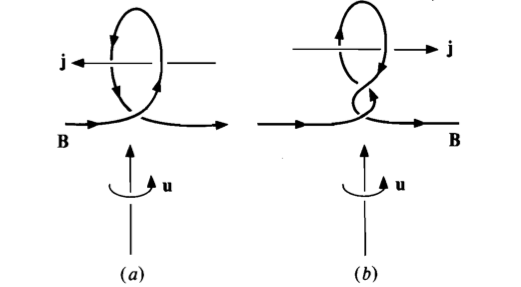}}
  \caption{ Field distortion by a localised helical disturbance. In ($a$) the loop is twisted through an angle $\pi/2$ and the associated current is anti-parallel to $\bfB$; in ($b$) the twist is $3\pi/2$, and the associated current is parallel to $\bfB$. (From \cite{Moffatt_1978}.)}
\label{fig:Parker_loops}
\end{figure}

A tacit assumption of expressions such as \eqref{eq:coeffs_ss} is that the averages are readily calculable, in the sense that the mean value of the e.m.f., averaged over the domain, is representative of a typical local value of the e.m.f. In this regard, it is instructive to reconsider Parker's original picture of rising, twisting and merging flux loops. As can be seen from Figure~\ref{fig:Parker_loops}, there will be a component of the induced current that is either anti-parallel or parallel to the untwisted initial field. As pointed out by \cite{Moffatt_1978}, if diffusion dominates or the `cyclonic events' are short lived, then the twist of the loops will be small, and the associated current of each loop will be anti-parallel to the field; therefore on averaging, all of the loops are acting in concert. This picture thus ties in nicely with the result~\eqref{eq:coeffs_low_Rm} (for low values of $Rm$) or \eqref{eq:coeffs_ss} (for short-sudden turbulence, $S \ll 1$). However, as discussed above, typical astrophysical turbulence does not fall into either of these regimes; persisting with the Parker picture, we might conclude that loops could be extremely twisted, with essentially a random distribution in the directions of each elemental current loop, and hence a small mean value.

\begin{figure}
  \centerline{\includegraphics[width=0.85\hsize]{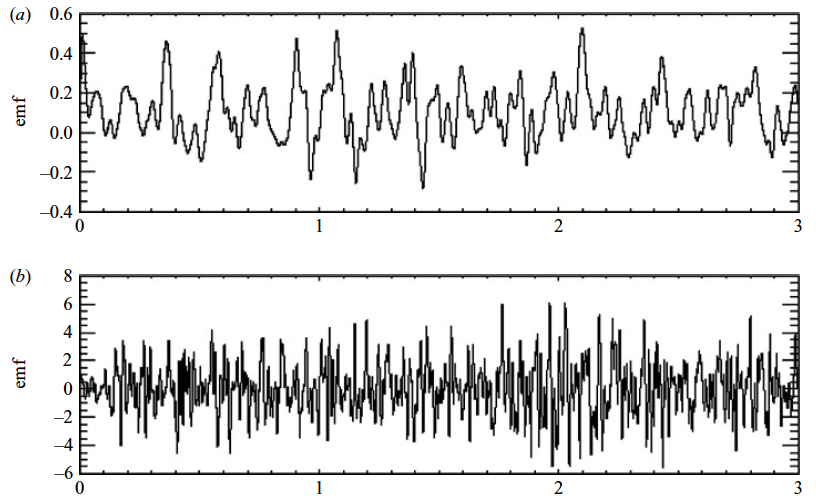}}
  \caption{Time histories of the longitudinal component of the e.m.f.\ for rotating plane layer convection, with aspect ratio $\lambda = 5$, imposed field strength $B_0 = 0.1$ and (a)~$Ra = 80,000$, (b)~$Ra=150,000$. (From \cite{HC_2008}.)}
\label{fig:CH_2008_Fig3}
\end{figure}

\begin{figure}
  \centerline{\includegraphics[width=0.85\hsize]{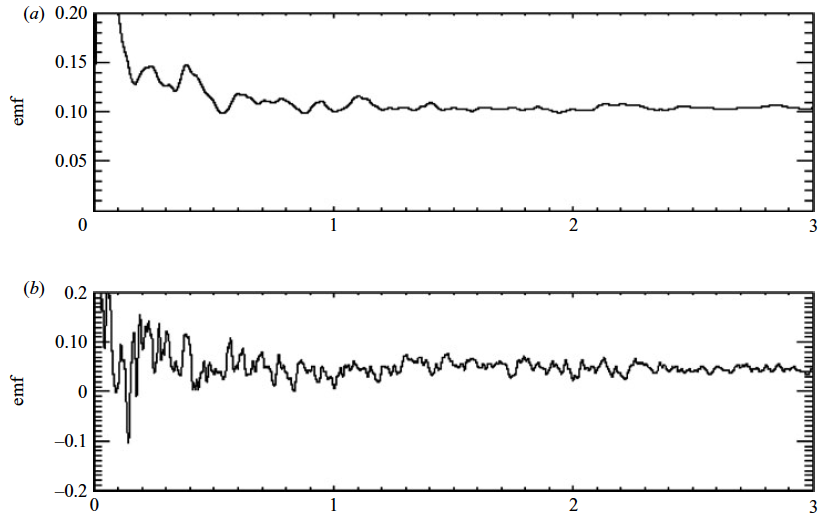}}
  \caption{Cumulative averages of the longitudinal component of the e.m.f.\ as a function of averaging length, for the cases illustrated in Figure~\ref{fig:CH_2008_Fig3}. (From \cite{HC_2008}.)}
\label{fig:CH_2008_Fig4}
\end{figure}

The issue of a potentially small value of the mean e.m.f.\ was addressed for rotating plane-layer convection by \cite{HC_2008}. They considered a regime in which the convection is turbulent, but still rotationally influenced, as revealed by the flow being noticeably helical. Figure~\ref{fig:CH_2008_Fig3} shows the time history of the longitudinal component of the spatially-averaged e.m.f.\ resulting from the imposition of a horizontal uniform field of strength $B_0=0.1$; the Taylor number is $Ta=500,000$, Prandtl number $Pr=1$ and magnetic Prandtl number $Pm=5$. Figure~\ref{fig:CH_2008_Fig4} shows the corresponding cumulative averages. The imposed field is very weak, with $B_0^2/\la \bfu^2 \ra \approx 2.8 \times 10^{-5}$ for $Ra = 80,000$ and $6.2  \times 10^{-6}$ for $Ra = 150,000$. At these parameter values the critical Rayleigh number for the onset of dynamo action is $Ra \approx 170,000$; for the two cases shown, the e.m.f.\ is thus solely a result of the imposed field.

The most striking aspect of Figure~\ref{fig:CH_2008_Fig3} is the wide variability in the e.m.f.\ with time, particularly bearing in mind that at each instant the plotted e.m.f.\ is the result of spatial averaging over a large domain (of size $5 \times 5 \times 1$) containing many turbulent eddies. At the higher Rayleigh number, the flow is more turbulent and the variability in the e.m.f.\ yet more pronounced. From Figure~\ref{fig:CH_2008_Fig4} it can be seen that after very long time averaging a clear value of the mean e.m.f.\ (and hence $\alpha$) eventually emerges. In the light of the preceding discussion, it is worth pointing out that the values of $\alpha$ are small; in particular they are much smaller than a characteristic velocity: for $Ra = 80,000$, $\alpha \approx 1$, whereas $\la \bfu^2 \ra^{1/2} \approx 20$; for $Ra = 150,000$, $\alpha \approx 0.4$, whereas $\la \bfu^2 \ra^{1/2} \approx 60$. For the case of $Ra=150,000$ the long-time average value of $\alpha$ differs from a typical value (already spatially averaged) by a factor of $10^2$, which is $O(Rm$); thus, as a result of the incoherence of the e.m.f., its mean magnitude is determined by diffusive rather than dynamic processes.

\subsection{Small-scale dynamo action}\label{subsec:ssda}

A potentially even more serious issue comes from re-examination of expression~\eqref{eq:fluct2}. The framework of mean field theory is built on the idea that the fluctuating field results solely from interactions between the small-scale velocity and the mean (large-scale) magnetic field; in other words, that all solutions of ${\cal L} (\bfb) = 0$ decay in time. When the theory of mean field electrodynamics was first formulated, this was a perfectly reasonable assumption, since very little was known about what we would now refer to as \textit{small-scale dynamos} (sometimes referred to as \textit{fluctuation dynamos}), in which the amplification of magnetic energy is entirely due to interactions between the small-scale flow and the small-scale field. Interest in, and understanding of, small-scale dynamos grew tremendously in the 1980s and 1990s with the detailed study of \textit{fast} dynamo action; i.e.\ (kinematic) dynamo action that persists in the limit as $Rm \to \infty$. This fascinating aspect of dynamo theory is described in detail in the monograph by \cite{CG_1995}. (Although fast dynamos are not necessarily small-scale, nearly all viable fast dynamo candidates are of this form.) Fast dynamo action is of a very different character to mean field generation: a lack of reflectional symmetry is no longer of central importance \citep[see][]{HCK_1996}; instead, what matters is whether, as discussed in the pioneering idea of \cite{VZ_1972}, the flow can stretch, twist and fold the magnetic field, causing its amplification in the limit of vanishing diffusivity; these ideas can be made quantitative through the notions of Lyapunov numbers and cancellation exponents \citep{DO_1993}.

What has become clear, from the computational study of small-scale dynamo action over the past thirty years or so, is that essentially all turbulent flows, at sufficiently high values of $Rm$, act as dynamos, i.e.\ as (exponential) amplifiers of the magnetic energy. However the fields are small-scale --- on the scale of the flow or smaller --- with no significant large-scale component. The important question then arises as to where does the idea of small-scale dynamo action fit into the theory of mean field electrodynamics. Partially to address this question we may consider the implications of small-scale dynamo action for the calculation of $\bfalpha$ by the prescription outlined in \S\,\ref{sec:cmft}. There are two issues to address: one is the practicable issue of calculation, the other is of interpretation; these are discussed in much more detail in \cite{CH_2009}.

If the flow acts as a small-scale dynamo then for a uniform imposed field $\bfB_0$, the fluctuating field, and hence the e.m.f., grows exponentially. Even if measurable, this gives an $\alpha$-effect that grows exponentially with time, which does not seem particularly helpful. In such a case one might instead consider $\bfB_0$ to be a large-scale component of the small-scale dynamo, in which case both $\bfcalE$ and $\bfB_0$ will grow exponentially at the same time. The interpretation of the tensors $\alpha$ and $\beta$ derived in this way is though certainly problematic, since, crucially, the growth rate of the observed field has nothing to do with that predicted by equation~\eqref{eq:grate}. A particularly striking example can be seen for the case of a non-helical dynamo. Here, $\alpha$ is zero and equation~\eqref{eq:grate} therefore predicts that large-scale averages should decay exponentially at a scale-dependent rate. However, as we have just argued, any average of the small-scale dynamo will grow exponentially at the small-scale dynamo growth rate, with this growth rate being determined not from mean field dynamo considerations, such as lack of reflectional symmetry, but from small-scale dynamo considerations, such as stretching and constructive folding. For a helical flow, a true mean field dynamo would emerge as the only candidate at low $Rm$; at higher $Rm$ this dynamo may still exist, but it would be unobserved since its growth would be swamped by the small-scale dynamo \citep[see][]{STP_2017}.

\section{The Dynamical Regime}\label{sec:DynReg}

Historically, mean field dynamo theory has resulted from consideration of the kinematic evolution of the magnetic field, i.e.\ that described by the induction equation for prescribed flows. However, in nature, there are no kinematic fields; the Lorentz force matters and the velocity and magnetic fields are related in a complex nonlinear fashion. It thus becomes important to consider how the mean field ideas of \S\,\ref{sec:KinReg} can be extended into the nonlinear regime. There are, broadly, two ways in which one might do this. One, discussed in \S\,\ref{sec:CHP}, which is closest in spirit to the original formulation, is to consider magnetohydrodynamic (as opposed to hydrodynamic) basic states and then to examine long wavelength instabilities of such states; this leads not only to extensions of existing mean field ideas, but also to new physical mechanisms, the implications of which have not yet been fully explored. The other is to adopt the mean field ansatz~\eqref{eq:emf}, together possibly with a similar mean-field formulation for the velocity, and to calculate the dependence of the mean field tensors on the strength of the mean magnetic field (and flow). 

\subsection{Mean field instabilities of MHD states}\label{sec:CHP}

A natural extension of classical mean field theory (i.e.\ the kinematic evolution of the field from a hydrodynamic state) is to consider mean-field instabilities, involving both the magnetic field and the flow, from a fully nonlinear magnetohydrodynamic basic state. For example, one might ask whether such instabilities can result from the saturated state of a small-scale dynamo, driven by some forcing $\bfF$. Somewhat surprisingly, these ideas have been explored only fairly recently \citep{CHP_2010a, CHP_2010b}.

In a pioneering paper of mean field electrodynamics, \cite{Roberts_1970} investigated the nature of the $\alpha$-effect by considering the instability of spatially periodic hydrodynamic flows to long wavelength perturbations in the magnetic field. \cite{CHP_2010a} extended this idea to consider long wavelength perturbations to fully magnetohydrodynamic flows; in such an approach, the velocity and magnetic field, in both the basic state and the perturbations, are of equal significance. In brief, the idea is as follows.

The starting point is a three-dimensional basic state $\bfB(\bfx,t)$, $\bfU(\bfx,t)$, which, for simplicity, we shall assume to be spatially and temporally periodic. We then consider long-wavelength perturbations, with wave vector $\bfk$, of the form
\begin{equation}
\left( \bfb ( \bfx, t), \bfu(\bfx,t) \right) = \left( {\bm H}(\bfx,t), {\bm V}(\bfx,t) \right)\, \exp \left( i \bfk \cdot \bfx + p(\bfk) t \right),
\label{eq:kansatz_2}
\end{equation}
where $\bfV$ and $\bfH$ vary on the same spatial and temporal scales as the basic state, and $p(\bfk)$ is the $\bfk$-dependent growth rate of the perturbations. The wavelength of the perturbations, $2 \pi / |\bfk|$, is assumed to be long in comparison with the spatial periodicity of the basic state. We now decompose the perturbations into average and fluctuating parts, with the average defined by
\begin{equation}
\left(\overline \bfb , \overline \bfu \right) = \left( \langle \bfH \rangle, \langle \bfV \rangle \right)\, \exp \left( {i\bfk\cdot\bfx +p(\bfk)t} \right),
\label{}
\end{equation}
where $\langle \ \rangle$ denotes an average over the spatial and temporal periodicities of the basic state; $\langle \bfH \rangle$ and $\langle \bfV \rangle$ are thus constants. The aim is to employ a mean field approach in order to determine the growth rate $p(\bfk)$ of modes with non-zero mean magnetic fields, i.e.\ modes for which $\langle \bfH \rangle \neq \textbf{0}$.

The magnitude of the wave vector $k = |\bfk|$ is treated as a small parameter; $\bfH$, $\bfV$ and the growth rate $p$ are then expanded in powers of $k$, so that, for example, 
\begin{equation}
\bfH = \bfH_0+\bfH_1+ \ldots \bfH_n +\ldots,
\label{eq:3D4}
\end{equation}
where $\bfH_n$ is of $n^{th}$ order in the components of $\bfk$.
At zeroth order, the governing equations become
\begin{align}
p_0\bfH_0 + \left(\partial_t - Rm^{-1}\nabla^2 \right) \bfH_0  &= \bmnabla \times \left( \bfV_0 \times \bfB + \bfU \times \bfH_0\right),
\label{eq:3D5} \\
p_0 \bfV_0 +\left( \partial_t - Re^{-1}\nabla^2 \right) \bfV_0 &=-\bmnabla \Pi_0 + \bmnabla \cdot \left( \bfH_0\bfB + \bfB\bfH_0 -\bfV_0\bfU - \bfU\bfV_0\right),
\label{eq:3D6} \\
\bmnabla \cdot \bfH_0 = 0, \quad
&\bmnabla \cdot \bfV_0 = 0, \label{eq:3D7}
\end{align}
where $\Pi$ denotes the perturbation to the pressure. Averaging equations~\eqref{eq:3D5} -- \eqref{eq:3D6} gives
\begin{equation}
p_0 \langle \bfH_0 \rangle = p_0\langle \bfV_0 \rangle = 0.
\label{eq:3D8}
\end{equation}
For non-zero mean field solutions we therefore need to take $p_0 = 0$. The corresponding equations for the fluctuations then read
\begin{align}
(\partial_t - Rm^{-1}\nabla^2 ) \bfh_0  &= 
\bmnabla\times \left( \bfv_0 \times \bfB + \bfU \times \bfh_0\right)+ \langle \bfH_0 \rangle \cdot \bmnabla \bfU -\langle \bfV_0 \rangle\cdot \bmnabla \bfB,
\label{eq:3D9} \\
( \partial_t - Re^{-1}\nabla^2 ) \bfv_0 &=-\bmnabla\Pi_0 + \bmnabla \cdot \left( \bfh_0\bfB + \bfB\bfh_0 -\bfv_0\bfU - \bfU\bfv_0\right)\nonumber \\
&\qquad \qquad \qquad \qquad \qquad + \langle \bfH_0 \rangle \cdot \bmnabla\bfB -\langle \bfV_0 \rangle\cdot\bmnabla\bfU,
\label{eq:3D10} \\
\bmnabla \cdot \bfh_0 = 0, \quad
&\bmnabla \cdot \bfv_0 = 0. \label{eq:3D11}
\end{align}
To first order, the equations for the mean variables give
\begin{align}
&p_1 \langle \bfH_0 \rangle = i\bfk\times \langle \bfv_0 \times \bfB + \bfU \times \bfh_0 \rangle, \label{eq:3D12} \\
&p_1 \langle \bfV_0 \rangle = -i\bfk \Pi_0 +i\bfk\cdot \langle \bfh_0 \bfB + \bfB\bfh_0 - \bfv_0\bfU -\bfU\bfv_0 \rangle, \label{eq:3D13} \\
&\bfk \cdot \langle\bfH_0 \rangle= 0, \quad
\bfk \cdot \langle\bfV_0 \rangle= 0. \label{eq:3D14} 
\end{align}
It can be seen from equations~(\ref{eq:3D9}) -- (\ref{eq:3D10}) that $\bfh_0$ and $\bfv_0$ are subject to a linear forcing by both $\langle \bfH_0 \rangle$ and $\langle \bfV_0 \rangle$. Therefore, expressions~(\ref{eq:3D12}) and (\ref{eq:3D13}) can be rewritten as
\begin{align}
&p_1 \langle \bfH_0 \rangle = i\bfk\times \left( \bfalpha^B \cdot\langle \bfH_0 \rangle + \bfalpha^U \cdot\langle \bfV_0 \rangle \right), 
\label{eq:3D15} \\
&p_1 \langle \bfV_0 \rangle = -i\bfk \Pi_0 +i\bfk\cdot \left( \bfGamma^B \cdot\langle \bfH_0 \rangle + \bfGamma^U \cdot\langle \bfV_0 \rangle \right). 
\label{eq:3D16} 
\end{align}
The evolution of any long wavelength instability to the magnetohydrodynamic basic state is thus governed by the four (constant) tensors $\bfalpha^{B}$, $\bfalpha^{U}$, $\bfGamma^{B}$ and $\bfGamma^{U}$. These depend on the basic state $\bfU$, $\bfB$, and hence on the forcing $\bfF$ and on the fluid and magnetic Reynolds numbers. For the kinematic dynamo problem, the only non-zero tensor is $\bfalpha^B$, which reverts to the standard $\alpha$-effect tensor. Similarly, for the mean field vortex instability --- in the absence of magnetic field --- the only non-zero tensor is $\bfGamma^U$, which describes the anisotropic kinetic alpha (AKA) effect of \cite{FSS_1987}. Finding a simple interpretation of these four tensors is even more difficult than for the purely kinematic problem, in which one has only the tensor $\bfalpha^B$. As in the kinematic case however, some progress can be made in the limit of small $Rm$ \citep{CHP_2010b}; under this restriction, the (dimensionless) e.m.f.\ can be approximated at leading order by
\begin{align}
\nonumber
\lambda^2 R_m^{-1} {\cal E}_i &\approx \bfU_0 \cdot \langle (1-P_m^{-1}) \epsilon_{ijk} \nabla u_j b_k \rangle \qquad  \qquad \\& \qquad \qquad \qquad + \bfB_0 \cdot \langle \epsilon_{ijk} ( u_j \nabla u_k + P_m^{-1} \chi \nabla b_j b_k) \rangle ,
\label{eq:indterm}
\end{align}
where $\lambda$ is the monochromatic wavenumber of the forcing (in both the momentum and induction equations), $\chi$ is a dimensionless parameter related to the amplitudes of the forcings and $Pm$ is the magnetic Prandtl number. A related expression can be found for the mean stress in the momentum equation. Even in this simplest case, in which we have an explicit expression for the e.m.f.\ in terms of the field and flow, it should be recalled that it is the forcing that is prescribed and that the field and flow follow as fully nonlinear states of the MHD equations. 

That said, it is of interest to note the form of the $\bfalpha^B$ term (i.e.\ the coefficient of $\bfB_0$) in expression~\eqref{eq:indterm}. It involves both the flow helicity $\langle \bfu \cdot {\bmnabla} \times \bfu \rangle$ and the current helicity $\mu_0^{-1} \langle \bfb \cdot {\bmnabla} \times \bfb \rangle$, and is reminiscent of the famous formula of \cite{PFL_1976}, namely
\begin{equation}
\alpha = - \frac{\tau_c}{3} \left( \la \bfu \cdot \bnabla \times \bfu \ra - \frac{1}{\mu_0 \rho} \la \bfb \cdot \bnabla \times \bfb \ra \right),
\label{eq:PFL}
\end{equation}
derived by applying the EDQNM approximation to homogeneous, isotropic, helical MHD turbulence. It is tempting to regard~\eqref{eq:PFL} as a formula describing how the nonlinear saturation of a dynamo driven kinematically by the flow helicity is achieved through the counteracting build up of current helicity. However, as pointed out by \cite{Proctor_2003}, this temptation should be resisted. If, for simplicity, we make the short-sudden approximation, then the expression~\eqref{eq:coeffs_ss} for $\alpha$, namely
\begin{equation}
\alpha = - \frac{\tau_c}{3} \la \bfu \cdot \bnabla \times \bfu \ra,
\label{eq:alpha_ss}
\end{equation}
follows immediately from the induction equation, \textit{whether the flow is kinematic or dynamic}. Thus all of the nonlinear effects are already included in \eqref{eq:alpha_ss}, through modification of the helicity or the correlation time (or both). In order to obtain an expression containing both kinetic and current helicities, it is instead necessary to start from a state of MHD turbulence, with a pre-existing flow $\bfu$ \textit{and} magnetic field $\bfb$. On then imposing a uniform field $\bfB_0$, the resulting perturbations to the flow and field can be approximated, at leading order, by
\begin{align}
\label{eq:ufluc}
\pt{\bfu'} &= -\bmnabla p +  \frac{1}{\mu_0 \rho} \bfB_0 \cdot \bmnabla \bfb, \\
\pt{\bfb'} &=  \bfB_0 \cdot \bmnabla \bfu.
\label{eq:bfluc}
\end{align}
The e.m.f.\ is given, to leading order, by
\begin{equation}
{\bfcalE} = \la \bfu \times \bfb' + \bfu' \times \bfb \ra,
\end{equation}
from which expression~\eqref{eq:PFL} follows, on solving~\eqref{eq:ufluc} and \eqref{eq:bfluc} for $\bfu'$ and $\bfb'$.

\begin{figure}
\centerline{\includegraphics[width=1.0\hsize]{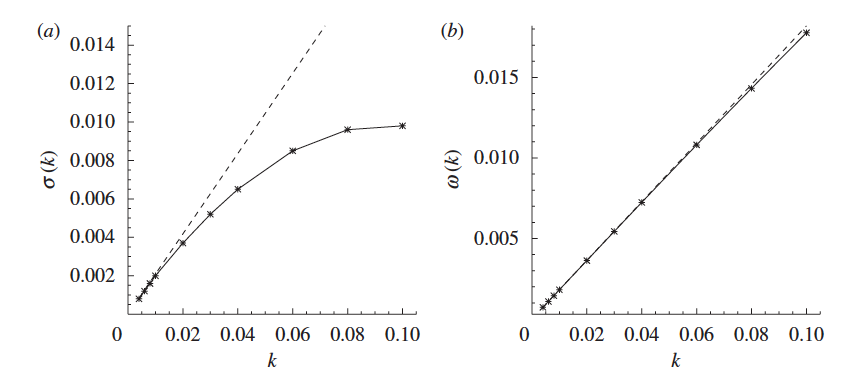}}
  \caption{($a$) Growth rate and ($b$) frequency of the instability of an MHD basic state, determined both by the mean field approach (dashed line) and by a solution of the full instability problem (solid line). The basic state results from applying the AKA forcing with an imposed magnetic field; $Rm=16$. (From \cite{CHP_2010a}.)}
\label{fig:CHP_fig2}
\end{figure}

Finally we should consider the practicalities of evaluating the four tensors $\bfalpha^B$, $\bfalpha^U$, $\bfGamma^B$ and $\bfGamma^U$. In theory, they can be calculated by solving equations~(\ref{eq:3D12}) -- (\ref{eq:3D14}). Sometimes, as in the standard determination of $\bfalpha$, everything works beautifully; similarly, sometimes it does not. \cite{CHP_2010a} extended the classic work of \cite{Roberts_1972} by considering two-dimensional magnetohydrodynamic states (dependent on $x$, $y$ and $t$), attained by applying a forcing to a basic state with an imposed uniform field (attaining a two-dimensional MHD state via saturation of a dynamo is of course impossible). They then evaluated the four governing tensors and determined the resulting growth rate of an instability (of both field and flow) with a long wavelength in the $z$ direction. Furthermore, they could also solve the full instability problem independently and compare the two approaches. Figure~\ref{fig:CHP_fig2} plots the real and imaginary parts of the growth rate, showing the comparison between the mean field approach and a full solution of the dynamo problem, for the case when the flow is forced by the AKA forcing of \cite{FSS_1987}, with $Rm=16$. Since the mean field theory is performed to first order in wavenumber $k$, the associated growth rate is linearly dependent on $k$; the agreement for small $k$ is excellent. 

In all of the examples consider by \cite{CHP_2010a}, it was straightforward to evaluate the governing tensors, with averages converging quickly. However, for three dimensional MHD states --- and indeed possibly for two-dimensional states at higher $Rm$ --- problems can arise in that linear perturbations to a state perturbed by a kinematic flow and field may have exponentially growing solutions; we then encounter a similar problem to that which undermines the determination of the kinematic $\alpha$-effect in the case when a small-scale dynamo is present. However, we may then ask the question as to what happens if instead of imposing a \textit{kinematic} field (or flow) we impose a \textit{dynamical} magnetic field, retaining the Lorentz force. Clearly, on energetic grounds, adding a very weak (but dynamic) magnetic field to an equilibrated MHD state will not lead to unbounded perturbations. The interesting question is whether, if the field is made sufficiently weak, there is an e.m.f.\ that is linear in the imposed field --- as assumed by the theory. In other words, can we salvage the linear theory in the circumstances for which the tensors are not calculable by considering the weak field limit of a dynamical field? This is clearly related to linear response theory, but the dependence of an averaged quantity to an external disturbance to a turbulent system is not necessarily straightforward \citep{HMPR_2018}.

\subsection{Nonlinear mean field tensors: the quenching problem}

A different approach to that outlined in \S\,\ref{sec:CHP} is to assume that, even in its dynamic state, the mean magnetic field still satisfies equation~\eqref{eq:mean_ind2}, but that all of the nonlinear effects are somehow incorporated into the $\bfalpha$ and $\bfbeta$ tensors. Attention has mainly focussed on the dependence of the $\alpha$-effect on the mean field $B_0$ --- what is known as the question of $\alpha$-\textit{suppression} or $\alpha$-\textit{quenching}. In dynamo models, one might also have a coupled mean field equation for the differential rotation, where the influence of the Reynolds stresses on the large scales results in what is known as a $\Lambda$-\textit{effect} \citep{Rudiger_1989, RH_2004}. One can then also incorporate $\Lambda$-quenching. Here, since we are concentrating on the fundamental aspects of the mean field formulation, rather than looking at specific dynamo models, we shall restrict attention just to the suppression of the transport terms in the induction equation.

The idea of incorporating a dynamical element into the $\alpha$-effect in order to prevent unlimited growth of the magnetic field is long standing \citep[see][]{Stix_1972, Jepps_1975}. From an astrophysical perspective, the  particular interest is in the nature of the suppression at very high values of $Rm$, and it is this issue that has led to some heated discussions over a number of years. At the heart of the matter is the question of how a magnetic field that is extremely weak on the large scales can, nonetheless, become dynamically significant on smaller scales. In a seminal paper, \citet{VR_1991} proposed a mechanism by which an extremely weak large-scale magnetic field could suppress turbulent magnetic diffusion. More specifically, they argued that even if the energy in the large-scale field were as small as the kinetic energy divided by $Rm$, then the associated small-scale field would be sufficiently strong to inhibit the turbulent diffusion process. These ideas were substantiated for the case of two-dimensional magnetohydrodynamic turbulence by the illuminating computations of \citet{CV_1991} and \cite{Cattaneo_1994}.

\begin{figure}
  \centerline{\includegraphics[width=0.7\hsize]{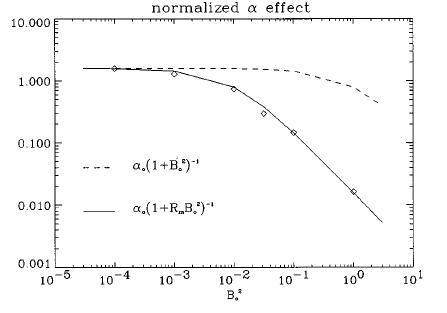}}
  \caption{Normalised $\alpha$-effect, $\alpha_N=\alpha_{33}/\langle \bfu^2 \rangle$, as a function of $B_0^2$ for cases with $Rm = Re = 100$. Each diamond corresponds to a numerical simulation. The dashed and solid curves are fits to the two different suppression formulae; in both cases the value at $B_0^2 = 10 ^{-4}$ has been fitted exactly. (From \cite{CH_1996}.)}
\label{fig:CH_1996}
\end{figure}

Further theoretical and numerical work undertaken in this direction by various authors \citep{KA_1992, TCV_1993, GD_1994, CH_1996} demonstrated that similar principles to those underlying the suppression of magnetic diffusion in two dimensions may be used to explain the marked suppression of the mean-field dynamo $\alpha$-effect in three dimensions.  Such a suppression, sometimes referred to as the `catastrophic' quenching of the $\alpha$-effect, presents a serious difficulty for the operation of any $\alpha$-effect dynamo at high $Rm$ in the nonlinear regime. \cite{CH_1996} addressed this problem via a numerical experiment designed specifically to measure the e.m.f.\ in a helical turbulent flow. They considered the three-dimensional basic state arising from the nonlinear saturation of the \cite{GP_1992} dynamo with $Re = Rm = 100$; the e.m.f.\ is then measured after the imposition of a uniform magnetic field $B_0 \bfzhat$. Comparison was made between two possible suppression formulae, namely
\begin{equation}
\alpha = \frac{\alpha_0}{1+B_0^2/ \langle \bfu^2 \rangle } \qquad \textrm{and} \qquad
\alpha = \frac{\alpha_0}{1+ Rm B_0^2/ \langle \bfu^2 \rangle },
\end{equation}
with the value $\alpha_0$ determined from fitting at the weakest field strength. In the former prescription, the magnetic field can reach equipartition strength before seriously influencing the $\alpha$-effect; by contrast, in the latter, only a very weak large-scale magnetic field (of order $Rm^{-1}$ times the kinetic energy) is needed to suppress $\alpha$. It can be seen clearly from Figure~\ref{fig:CH_1996} that the numerical results provide an excellent agreement with the more dramatic suppression formula. Clearly, if the $\alpha$-effect is the dominant dynamo process in astrophysical contexts, in which $Rm$ is invariably huge, then this dramatic suppression presents a major problem in explaining observed large-scale magnetic fields. It is though worth emphasising that the issue is one of suppression of the \textit{turbulent} $\alpha$-effect, i.e.\ in which $\alpha_0 \sim \langle \bfu^2 \rangle^{1/2}$. In other words, an $\alpha$-effect that acts on a fast (flow) timescale will be suppressed on that timescale. On a sufficiently long (Ohmic) timescale (though astrophysically this is extremely long), such suppression presents no impediment to the build up of strong large-scale fields.

Despite strong numerical evidence, such as presented in Figure~\ref{fig:CH_1996}, together with underpinning theoretical arguments, the nature of the suppression of the $\alpha$-effect has proven to be a very controversial issue. The strong suppression result of \cite{CH_1996} was ascribed by \cite{BF_2000} as being due to the adoption of periodic boundary conditions, and, in particular, to their influence on the magnetic helicity. Indeed, they go so far as to say that `this (the strong suppression result) is not a dynamical suppression from the back-reaction but a constraint on the magnitude of $\alpha$ that is imposed by the boundary conditions'. This though is simply wrong --- whatever else one might argue about, the suppression shown in Figure~\ref{fig:CH_1996} is manifestly a consequence of the Lorentz force! However, that is not to say
that the influence of boundary conditions is not of interest. It certainly is, but numerical experiments carefully designed to explore this issue have not yet been performed.

Discussions on the saturation of the $\alpha$-effect, together with the influence of boundaries, are often couched in terms of the magnetic helicity, defined by
\begin{equation}
{\cal H}_M = \int \bfA \cdot \bfB \, \dV,
\label{eq:MagHel}
\end{equation}
where $\bfA$ is the magnetic vector potential, and are based on the twin notions that magnetic helicity is `almost conserved' and that dynamos can be best understood through consideration of the \textit{flux} of magnetic helicity. The idea that magnetic helicity is `almost conserved' was introduced in an extremely important paper by \cite{Taylor_1974}, with the very specific aim of explaining the generation of reversed fields in toroidal plasmas. This conjecture by Taylor does not of course hold in all circumstances (nor indeed did Taylor claim that it would) but, nonetheless, it seems to have become part of dynamo folklore, leading to the idea that large-scale fields can prosper only through a flux of large-scale helicity \cite[see, for example,][]{VC_2001}. Both of these ideas are though somewhat troublesome. Magnetic helicity is not necessarily conserved, or indeed `almost conserved', in a resistive fluid --- certainly during any kinematic dynamo phase, the magnetic helicity, unless identically zero, grows at precisely the same rate as the magnetic energy. In addition, magnetic helicity, as in \eqref{eq:MagHel}, is gauge invariant only when the volume $V$ is bounded by a flux surface; the notion of a \textit{flux} of magnetic helicity, a gauge-dependent quantity, from one scale to another, or through a boundary, is thus problematic. Although the notion of magnetic helicity can be extremely valuable in analysing the topology of magnetic field configurations \cite[see, for example,][]{Berger_1999}, it is worth bearing in mind, particularly in the dynamo context, that it is not something over and above the magnetic induction equation; from a solution of the induction equation, the magnetic helicity necessarily follows.

\section{Discussion}

After a whistle-stop tour through the ups and downs of some fundamental aspects of the theory of mean field electrodynamics, we should reflect on where we stand --- not just with the theory itself, but with the greater goal of explaining the maintenance of global-scale astrophysical magnetic fields. Somewhat oversimplifying matters, we may arrive at two very different viewpoints. On the one hand, it is certainly true that mean field models --- none of which I have discussed in this review --- are able to reproduce astrophysical magnetic fields extremely well; for example, it is possible to construct a mean field model, suitably parameterised, that has roughly periodic dynamo waves, propagating towards the equator, interspersed with intervals of greatly reduced activity --- very much like the solar dynamo. Optimistically, one might therefore argue that, at heart, a mean field approach is the right one for understanding astrophysical dynamos. Less optimistically, however, one might argue that the success of such models is due to the freedom allowed in the choice of $\alpha_{ij}$, $\beta_{ijk}$, etc.\ and that it is still extremely difficult to reconcile such choices with a more `bottom-up' approach to evaluating these transport tensors.

The most elemental problem we can consider in a mean field context is the kinematic growth of magnetic fields in a helical, turbulent flow at high $Rm$. In my view, it seems that small-scale dynamo considerations (i.e.\ stretching, twisting and folding) will dominate over any mean field characteristics (such as helicity) and that any large-scale component of the field will simply be as part of the small-scale eigenfunction. As for possible mean field instabilities of a fully MHD basic state, this problem has not yet been fully explored. Pessimistically, there is the worry that any predicted mean field instability, deduced from an argument relying on separation of spatial scales, will not manifest itself in a system that has many intermediate degrees of freedom --- rather than a long wavelength instability slowly emerging, as the theory would predict, it is conceivable that the basic state would simply adjust slightly to accommodate any perturbation.

Although maybe it is asking too much for large-scale fields to emerge solely from small-scale interactions (what would be termed an $\alpha^2$-dynamo), this requirement is anyway unrealistic in astrophysical bodies, which, typically, have a global-scale flow. In such an $\alpha \omega$-dynamo, the toroidal field arises from the shearing of the poloidal component by the differential rotation $\omega$; the $\alpha$-effect still has to close the dynamo loop by regenerating the poloidal field. Given some of the difficulties with $\alpha$, discussed above, we should consider how these might be allayed through the added ingredient of a global-scale shear flow or differential rotation. Within the mean field framework, one  can envisage various possible beneficial effects of a velocity shear on the mean field dynamo process \citep[see][]{HP_2009}: for example, (i)~that even though $\alpha$ is small, $\omega$ is so effective that there may nonetheless be an effective $\alpha \omega$-dynamo; (ii)~more subtly, that the large spatial scale of the shear leads to an enhanced $\alpha$ through greater spatial correlation of the small-scale motions;  (iii) that the anisotropy induced by the shear may lead to more exotic mean field effects (such as the shear-current effect of \cite{RK_2003}). Alternatively it may be that the answer lies outside the strictures of mean field theory.

Recently there have been various, rather different, studies to address this question through investigations of kinematic dynamo action driven by small-scale turbulence and a large-scale (imposed) shear flow. \cite{Yousef_etal_2008} showed how the combination of a shear flow with small-scale non-helical turbulence can lead to a large-scale magnetic field; here there is no $\alpha$-effect and so the dynamo mechanism certainly lies outside the standard $\alpha \omega$ picture. \cite{HP_2013} extended the plane layer rotating convection dynamo model of \cite{CH_2006} by imposing a large-scale uni-directional horizontal shear flow, dependent only on the other horizontal component; hydrodynamic interactions between the convection and the shear led to a flow with a broad range of scales. The presence of velocity shear both enhanced the dynamo growth rate and also led to the generation of significant magnetic field on the larger scales. However, by analysing spectrally filtered flows, it was shown that the dynamo depends crucially for its existence on the entire range of velocity scales, and that it could not be described in terms of a dynamo with scale separation. A somewhat different conclusion was reached by \cite{CT_2014}, who considered dynamo action driven by a combination of Galloway-Proctor cells (described by \eqref{eq:flows}), together with a much larger-scale steady shear flow. Here it was found that the shear could suppress the small-scale dynamo action, thereby allowing mean field processes to flourish, with the emergence of dynamo waves.

Thus the picture is still unclear, but it seems that it is through understanding the complex interactions between small-scale turbulence, large-scale flows and the magnetic field, together possibly with the influence of other large-scale inhomogeneities, that further progress will be made. Whether the answer will be as elegant as classical mean field theory remains to be seen. Fifty years on, there is still some way to go!

\medskip

I am grateful to STFC, who for many years have funded much of my own research on astrophysical dynamo theory.

\bibliographystyle{jpp}

\bibliography{MFE50}

\end{document}